\documentclass[twocolumn]{aastex62}
\usepackage{amssymb,bm,dcolumn,booktabs,threeparttable,multirow,fleqn}
\usepackage{amsmath}
\usepackage{multirow,bigdelim}
\usepackage{txfonts}
\usepackage{natbib}

\begin{document}

\title{Bayesian Analysis on the X-ray Spectra of the Binary Neutron
Star Merger GW170817}

%% Note that the corresponding author command and emails has to come
%% before everything else. Also place all the emails in the \email
%% command instead of using multiple \email calls.

%\correspondingauthor{En-Tzu Lin}

\author[0000-0002-0030-8051]{En-Tzu Lin}
\affiliation{Institute of Astronomy, National Tsing Hua University, Hsinchu 30013, Taiwan}
\email{entzulin@gapp.nthu.edu.tw}
\author[0000-0001-5643-7445]{Hoi-Fung Yu}
\affiliation{Faculty of Science, The University of Hong Kong, Pokfulam, Hong Kong}
\affiliation{Department of Physics, KTH Royal Institute of Technology, 10691 Stockholm, Sweden}
\affiliation{Oskar Klein Centre for Cosmoparticle Physics, 10691 Stockholm, Sweden}
\email{davidyu@hku.hk}
\author[0000-0002-5105-344X]{Albert K.H. Kong}
\affiliation{Institute of Astronomy, National Tsing Hua University, Hsinchu 30013, Taiwan}
\email{akong@gapp.nthu.edu.tw}

\keywords{gamma-ray burst: individual (GRB170817A)  --- gravitational waves --- X-rays: individual (GW170817)}

\begin{abstract}
For the first time, we present a Bayesian time-resolved spectral study of the X-ray afterglow datasets of GW170817/GRB17017A observed by the {\it Chandra} X-ray Observatory. These include all 12 public datasets, from the earliest observation taken at $t \sim 9$~d to the newest observation at $ \sim 359$~d post-merger. 
While our results are consistent with the other works using Cash statistic within uncertainty, the Bayesian analysis we performed in this work have yielded Gaussian-like parameter distributions. We also obtained the parameter uncertainties directly from their posterior probability distributions.
We are able to confirm that the power-law photon index has remained constant of $\Gamma \sim 1.6$ throughout the entire year-long observing period, except for the first dataset observed at $t = 8.9$~d when $\Gamma =1.04\pm0.44$ is marginally harder. We also found that the unabsorbed X-ray flux peaked at $t \sim 155$~d, temporally consistent with the X-ray flare model suggested recently by Piro et al (2018). The X-ray flux has been fading since $\sim160$ days after the merger and has returned to the level as first discovered after one year. Our result shows that the X-ray spectrum of GW170817/GRB170817A is well-described by a simple power-law originated from non-thermal slow-cooling synchrotron radiation.
\end{abstract}

%\titlerunning{A Possible X-ray Spectral Change from GW170817}
%\authorrunning{Lin et al.}
%\maketitle

\section{Introduction} \label{sec:intro} 
On 17 August 2017, a binary neutron star merging event triggered 
the first confirmed detection of gravitational wave (GW) signal known as GW170817 \citep{Abbott2017} 
accompanied by gamma-ray emission \citep{Goldstein2017,Savchenko2017}. 
On-going monitoring to this source has then revealed that the emissions 
span the whole electromagnetic spectrum. The X-ray counterpart of 
GW170817 was discovered by the \textit{Chandra} X-ray Observatory 9 days  
after the detection of the GW signal \citep{Troja2017}.

Binary neutron star mergers are thought to be the progenitors of short 
gamma-ray bursts (sGRBs). The sGRB signal associated with GW170817, named GRB170817A, has confirmed this long-standing hypothesis. However, the observational 
properties of GRB170817A is not similar to the majority of sGRBs across the 
electromagnetic spectrum, suggesting that GRB170817A is probably observed off-axis and structured \citep[e.g.,][]{Alexander2017,Evans2017,Fong2017,Haggard2017,Hallinan2017,Margutti2017,Troja2017}. Various models of GRB jet could be distinguished by the afterglow 
emission mechanism inferred from their spectral shapes.
This motivates us to reanalyze all the existing X-ray spectra of GRB170817A with Bayesian statistics.
%Thus, X-ray spectral study of GRB170817A is of great interest and importance. 

In the fireball model of GRBs \citep{Goodman1986,Paczynski1986,Rees1992,Piran1999}, materials are accelerated to relativistic speeds in the jet. Ejecta collide with the circumburst medium and the electrons within are shock-accelerated to a power-law population $N_\gamma \varpropto \gamma^{-p}$ with a minimum injection 
energy $\gamma_{\rm min}$ and cutoff energy $\gamma_{\rm cool}$, above which the 
electron cools significantly via synchrotron emission. This scenario, called the 
external shock model, results in a spectral shape with several power-law 
segments \citep[e.g.,][]{Sari1998,Granot2002}, 
\begin{equation}
F_\nu \varpropto \nu^{-\beta},
\label{eqn:powerlaw}
\end{equation}
where $\rm \beta$ is the spectral index such that $\rm \Gamma = \beta + 1$ is the 
photon index in the convolved photon spectrum. The observed value of $\Gamma$ depends on the observing energy band, evolution of the electron population, and the micro-physics in the ejecta. 
The relative position of the characteristic frequencies, $\nu_{\rm cool}$ and $\nu_{\rm min}$, changes with the value of $\Gamma$ (Eqns.~(\ref{eqn:slow}) and (\ref{eqn:fast})). These characteristic frequencies depend on the electron and magnetic equipartition factors, $\epsilon_{\rm e}$ and $\epsilon_{\rm B}$ \citep{Sari1998,Panaitescu2000}. Moreover, the value of the peak energy of the $\nu F_\nu$ spectrum also helps to constrain $\epsilon_{\rm e}$ and $\epsilon_{\rm B}$ \citep{Dermer1999}. 
Constraining the value of $\Gamma$ can therefore lead to an understanding of the emission processes taking place in the ejecta.

Here, we reanalyzed all archival X-ray spectral data of GRB170817A using Bayesian inference, which is a Bayesian statistical analysis technique mainly used in model parameter estimation. 
The conventional frequentist statistical method of using various test-statistics (e.g., $\chi^2$ minimization and its variants) might suffer from low number of photon counts \citep[see, Fig.~2 of][]{Greiner2016}. For instance, when the photon count is low ($\sim 20$ photons), it follows a Poisson statistic 
such that the Gaussian assumption of $\chi^2$ minimization may no longer hold. In such cases, the Cash statistic \citep{Cash1979} should be used instead. However, there remains the issue of estimating the parameter uncertainty (i.e., the error bar). Since these are point estimations, error bars are not given for free as in Bayesian approach and are always estimated around the mode (i.e., the maximum) of the test-statistic.
Bayesian inference can derive the posterior distributions of each parameter so that the errors (known as the credible regions) are readily obtained, while the distribution can be skewed or even multi-modal. 
Since individual time-resolved spectra of GRB170817A shows low signal-to-noise ratio, we therefore employ Bayesian inference in this paper to obtain the measurements and uncertainties of $\Gamma$ for individual time-resolved spectra from the public \textit{Chandra} X-ray datasets of GW170817/GRB170817A since its first X-ray detection.

This paper is arranged as follows. We present the data analysis method and results in Sect.~\ref{sec:meth}. We discuss our results in 
Sect.~\ref{sec:diss} and briefly summarize in Sect.~\ref{sec:con}. 
Unless otherwise stated, all uncertainties are given at the 1-$\sigma$ Bayesian credible level.

\section{Data Analysis and Results} \label{sec:meth}

\begin{table*}[htp]
\centering
\caption{ {\it Chandra} X-ray observations and resulting parameters and uncertainties.}
\label{tab:results}
\begin{tabular}{ccccccccc}
\hline
\hline
Episode & Obs.~ID & Exposure & Time since GW & S/N & $\Gamma^a$ & $\Gamma^b$ & $\Gamma({\rm ref.})^c$ & Flux (0.3-8.0 keV) \\
 & & (ks) & (days) & & & & &($10^{-15}{\rm erg~cm}^{-2}~{\rm s}^{-1}$) \\
\hline
\multirow{4}{*}{\uppercase\expandafter{\romannumeral 1}} & \multirow{2}{*}{19294} & \multirow{2}{*}{49.4} & \multirow{2}{*}{8.9} & \multirow{2}{*}{3.33} & \multirow{2}{*}{$1.04^{+0.44}_{-0.44}$} & \multirow{2}{*}{$2.2^{+1.6}_{-1.3}$} & ($0.95^{+0.95}_{-0.19}$)$^d$ & \multirow{2}{*}{$6.89^{+2.08}_{-2.34}$} \\
& & & & & & & ($0.9^{+0.5}_{-0.5}$)$^e$ & \\
 & \multirow{2}{*}{20728} & \multirow{2}{*}{46.7} & \multirow{2}{*}{15.2} & \multirow{2}{*}{2.85} & \multirow{2}{*}{$2.02^{+0.61}_{-0.49}$} & \multirow{2}{*}{$2.2^{+1.2}_{-1.5}$} & $1.6^{+1.5}_{-0.1}$)$^d$ & \multirow{2}{*}{$5.61^{+1.92}_{-1.50}$} \\
& & & & & & & ($2.42^{+0.95}_{-0.88}$)$^f$ & \\
\hline
\multirow{2}{*}{\uppercase\expandafter{\romannumeral 2}} & 20860 & 74.1 & 107.5 & 9.61 & $1.47^{+0.17}_{-0.16}$ & $1.3^{+0.15}_{+0.15}$ & \rdelim\}{2}{6pt}[]($1.62^{+0.16}_{-0.16}$)$^d$ & $23.1^{+2.26}_{-2.38}$ \\
 & 20861 & 24.7 & 110.9 & 6.12 & $1.91^{+0.34}_{-0.31}$ & $1.8^{+1.4}_{-1.3}$ & ~~($1.53^{+0.24}_{-0.23}$)$^f$ & $21.6^{+3.99}_{-4.66}$ \\
\hline
\multirow{5}{*}{\uppercase\expandafter{\romannumeral3}} & 20936 & 32.0 & 153.5 & 8.89 & $1.45^{+0.25}_{-0.24}$ & $2.0^{+0.83}_{-0.53}$ & \rdelim\}{5}{6pt}[]\multirow{5}{*}{\shortstack{($1.61^{+0.17}_{-0.17}$)$^d$\\($1.58^{+0.23}_{-0.22}$)$^f$}} & $29.0^{+4.69}_{-4.23}$ \\
 & 20937 & 20.7 & 156.4 & 9.78 & $1.77^{+0.35}_{-0.35}$ & $1.8^{+0.78}_{-0.75}$ & & $19.4^{+6.21}_{-4.09}$ \\
 & 20938 & 16.0 & 157.1 & 9.74 & $1.77^{+0.34}_{-0.35}$ & $1.9^{+0.75}_{-0.96}$ & & $27.9^{+6.24}_{-5.08}$ \\
 & 20939 & 22.5 & 159.9 & 4.41 & $2.03^{+0.44}_{-0.39}$ & $2.2^{+1.2}_{-1.0}$ & & $16.1^{+4.05}_{-3.68}$ \\
 & 20945 & 14.4 & 164.0 & 6.88& $2.00^{+0.48}_{-0.46}$ & $2.3^{+1.1}_{-1.2}$ & & $18.6^{+4.65}_{-5.06}$ \\
\hline
\multirow{3}{*}{\uppercase\expandafter{\romannumeral 4}} & 21080 & 50.7 & 259.4 & 4.20 & $1.66^{+0.32}_{-0.30}$ & $2.1^{+1.5}_{-1.1}$ & \rdelim\}{2}{6pt}[]\multirow{2}{*}{($1.57^{+0.38}_{-0.39}$)$^f$} & $11.5^{+2.40}_{-2.12}$ \\
 & 21090 & 46.0 & 261.0 & 4.67 & $1.63^{+0.32}_{-0.32}$ & $2.1^{+1.4}_{-1.1}$ & & $13.0^{+2.57}_{-2.23}$ \\
 & 21371 & 67.2 & 358.6 & 3.30 & $1.76^{+0.34}_{-0.34}$ & $2.2^{+1.4}_{-1.1}$ & $(1.6^{+1.3}_{-0.9})^{g}$ & $6.51^{+1.30}_{-1.17}$ \\
\hline
\end{tabular}
\begin{tablenotes}
\footnotesize
\item NOTE. All uncertainties are given at the 1-$\sigma$ Bayesian credible level.
\item $^a$ Derived from Sherpa.
\item $^b$ Derived from 3ML.
\item $^c$Reference values of $\Gamma$. Big right bracket means time-integrated spectrum.
\item $^d$From \citet{Margutti2018}.
\item $^e$From \citet{Troja2018}.
\item $^f$From \citet{Nynka2018}.
\item $^g$From \citet{Haggard2018}
\end{tablenotes}
\end{table*}

We employed Bayesian inference to analyze all existing public \textit{Chandra} datasets (Table.~\ref{tab:results}) observed from 26 August 2017 to 10 August 2018 (a total of 12 observations). 
Each dataset was processed with the \texttt{CIAO} software package (v4.9).\footnote{\url{http://cxc.harvard.edu/ciao/}}
The centroid position of the X-ray counterpart of GW170817/GRB170817A was determined by the point-source detection algorithm \texttt{wavdetect}.
We selected a $2 \farcs 0$ extraction region for the source, corresponding to a $\approx 90 \%$ of the encircled energy fraction, while the background was estimated from a nearby source-free region. The spectra and associated response 
files were obtained using \texttt{specextract}. We performed Bayesian spectral analysis with {\tt Sherpa} which is the modeling and fitting application of {\tt CIAO}. A power law in the form of Eqn.~(\ref{eqn:powerlaw}) 
with an absorption was fit to each dataset. {We fixed the absorption column density at $N_{H} = 7.5 \times 10^{20}~{\rm cm}^{-2}$, obtained by converting the Galactic optical extinction $A_{V}= 0.338$ \citep{Schlafly&Finkbeiner2011} to a hydrogen column density according to the relation $N_{H}(cm^{-2}) \approx 2.21 \times 10^{21} A_{V}$ \citep{Guver2009}.} Uniform priors for the normalization 
$K \sim \mathcal{U}(10^{-10}, 0.01)$ and photon index $\Gamma \sim \mathcal{U}(0.01, 5.0)$ were used to reflect our ignorance of the intrinsic distribution. In addition, we obtained consistent results using Gaussian priors. Conservatively, the results shown in this paper were adopted from uniform priors.
A Poisson likelihood was used for both the source and background spectra. 
We sampled the posterior distribution using Markov chain Monte Carlo (MCMC) sampling technique. The means and credible intervals of the posterior distributions are then computed. Note that the Bayesian implementation of {\tt Sherpa} took the background, 
instrumental response, and telescope's effective area as well as their 
uncertainties into account when sampling the posterior distribution \citep{vanDyk2001}.

\begin{figure}[htp]
\centering
\includegraphics[width=1.0\linewidth]{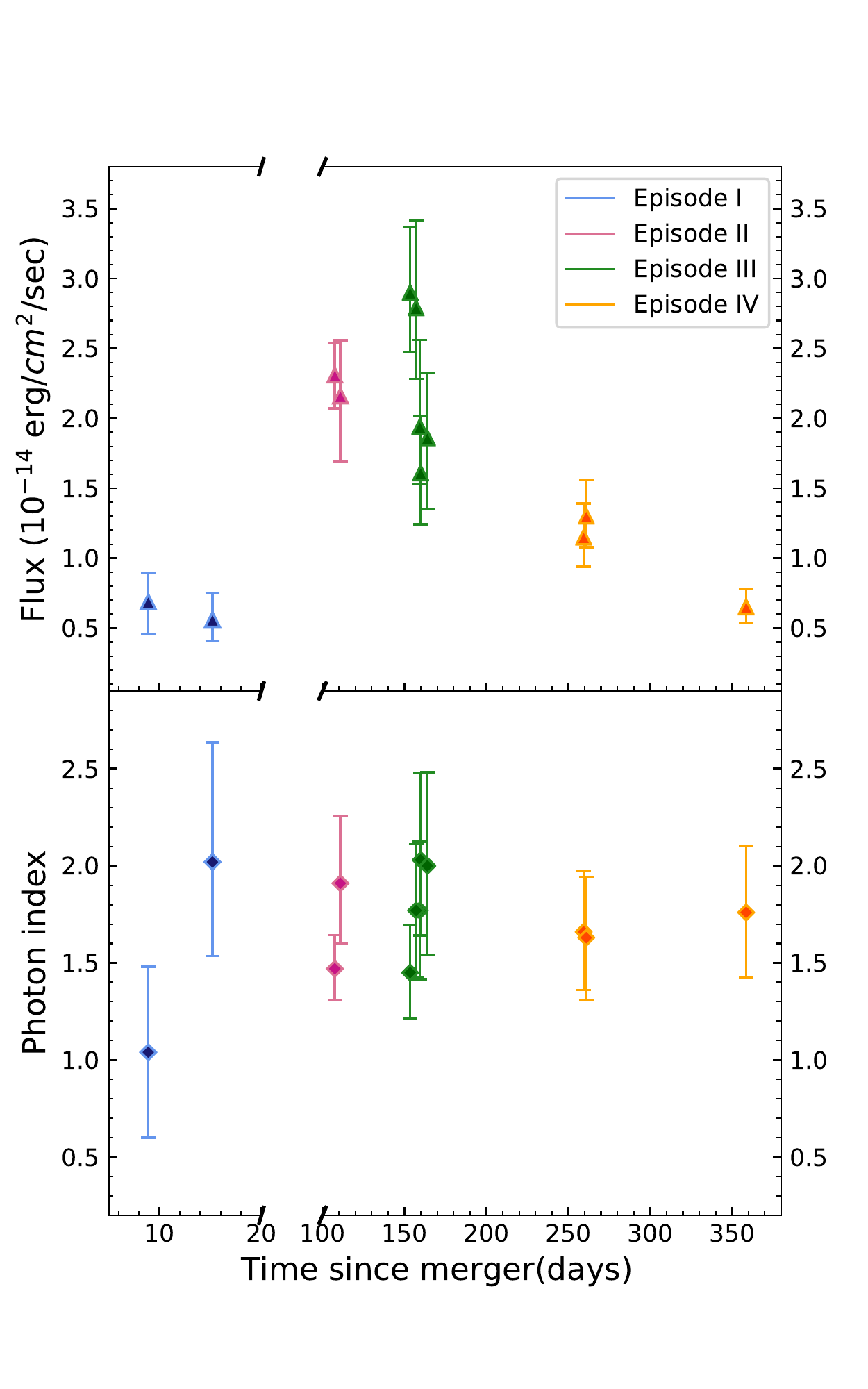}
\caption{Temporal evolution of the derived 0.3-8 keV unabsorbed flux (upper panel) and photon index $\Gamma$ (lower panel) listed in Table~\ref{tab:results}. Time is relative to the detection of the GW event.
\label{fig:1}}
\end{figure}

\begin{figure}[htp]
\centering
\includegraphics[width=1.00\linewidth]{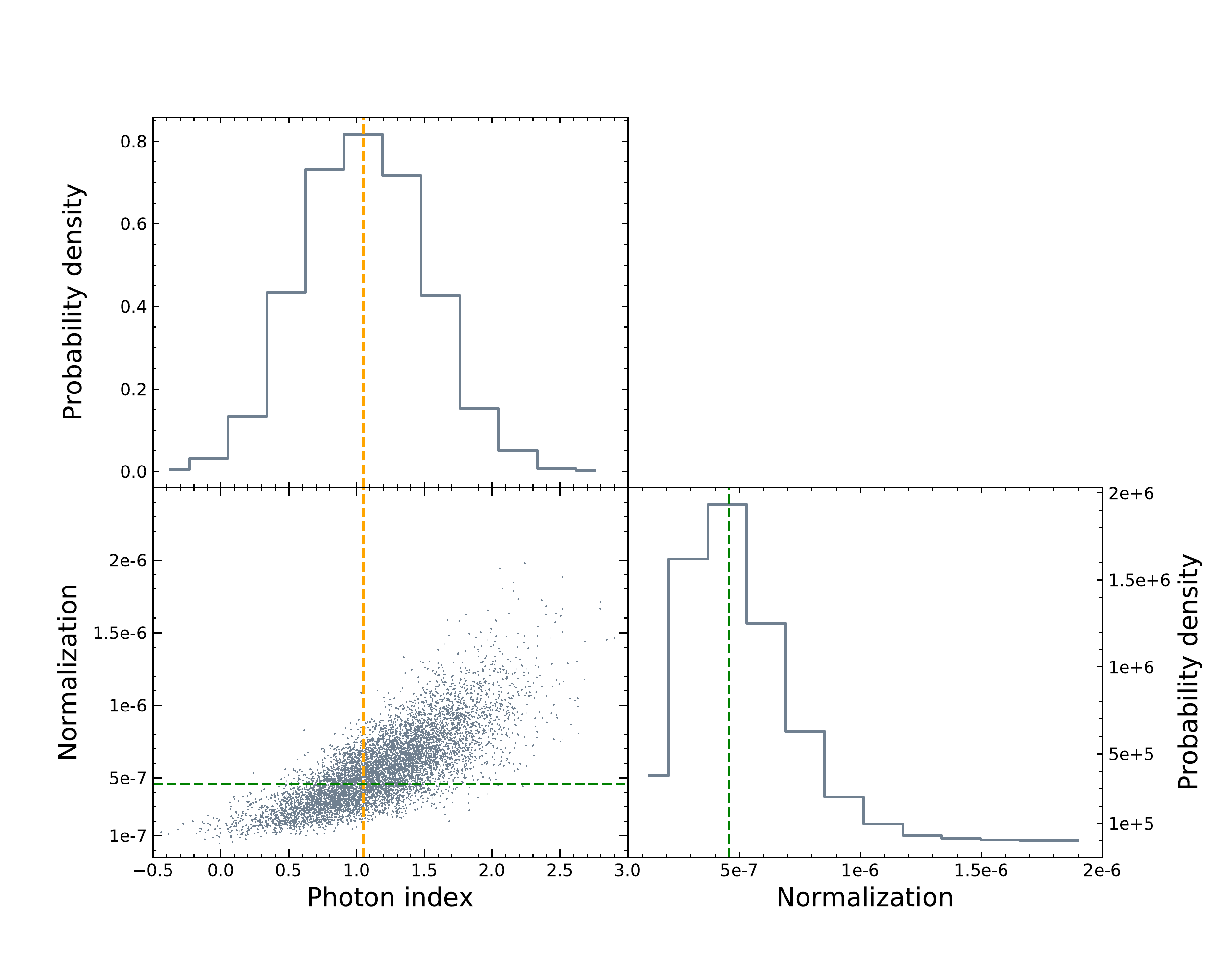}
\caption{Corner plot for the Bayesian inference of the first observation (ObsID 19294) 
showing the MCMC sampling (grey dots). The orange and green dashed lines show the 
median of $\Gamma$ and normalization, respectively. The histograms display the 
marginal distributions of the parameters.
\label{fig:2}}
\end{figure}

The 12 {\it Chandra} X-ray observations and Bayesian inference results are summarized in Table~\ref{tab:results} and plotted in Fig.~\ref{fig:1}. The corner plot of the posterior distribution of the first observation is illustrated in Fig.~\ref{fig:2}. 
According to the proximity of the observing time, the 12 observations can be divided into four episodes: (I) 8.9-15.2 days; (II) 107.5-110.9 days; (III) 153.5-164.0 days; and (IV) 259.4-358.6 days. 
We found an overall photon index ranging between $\Gamma = 1.47^{+0.17}_{-0.16}$ to $\Gamma=2.03^{+0.44}_{-0.39}$ (except the first dataset discussed below), consistent with a value of $\Gamma \sim 1.6$ from previous analysis \citep{Margutti2017,Mooley2018,Troja2017,Haggard2017,Margutti2018,Ruan2018,Troja2018,Nynka2018}, which is consistent with an election power-law distribution index $p=2.2$. The absorbed flux in 0.3-8.0~keV derived from the first X-ray counterpart detection at $t = 8.9$~d post-merger was $6.89 \times 10^{-15}~{\rm erg~cm}^{-2}~{\rm s}^{-1}$, and it gradually rose to a value of $2.31 \times 10^{-14}~{\rm erg~cm}^{-2}~{\rm s}^{-1}$ in episode II. Subsequent observations in episode III then revealed that the X-ray afterglow had reached its maximum intensity, of flux = $2.9 \times 10^{-14}~{\rm erg~cm}^{-2}~{\rm s}^{-1}$  at $t\sim 153.5$~d that was more than four times brighter than the first detection. After the peak, the flux swiftly decreased to $1.61 \times 10^{-14}~{\rm erg~cm}^{-2}~{\rm s}^{-1}$ within 6 days. Same temporal variation was also noticed by \citet{Piro2018}, in which they proposed an X-ray flare from the source was possibly observed. Following the fading trend seen in episode III, the source intensity continued to decrease in the late times. {The observed flux at $t \sim 359$~d was $6.51 \times 10^{-15}~{\rm erg~cm}^{-2}~{\rm s}^{-1}$. Almost one year after the merging event, the signal from the X-ray counterpart of GW170817 has now dropped to a similar intensity as first discovered.}

\begin{figure}[htp]
\centering
\includegraphics[width=1.00\linewidth]{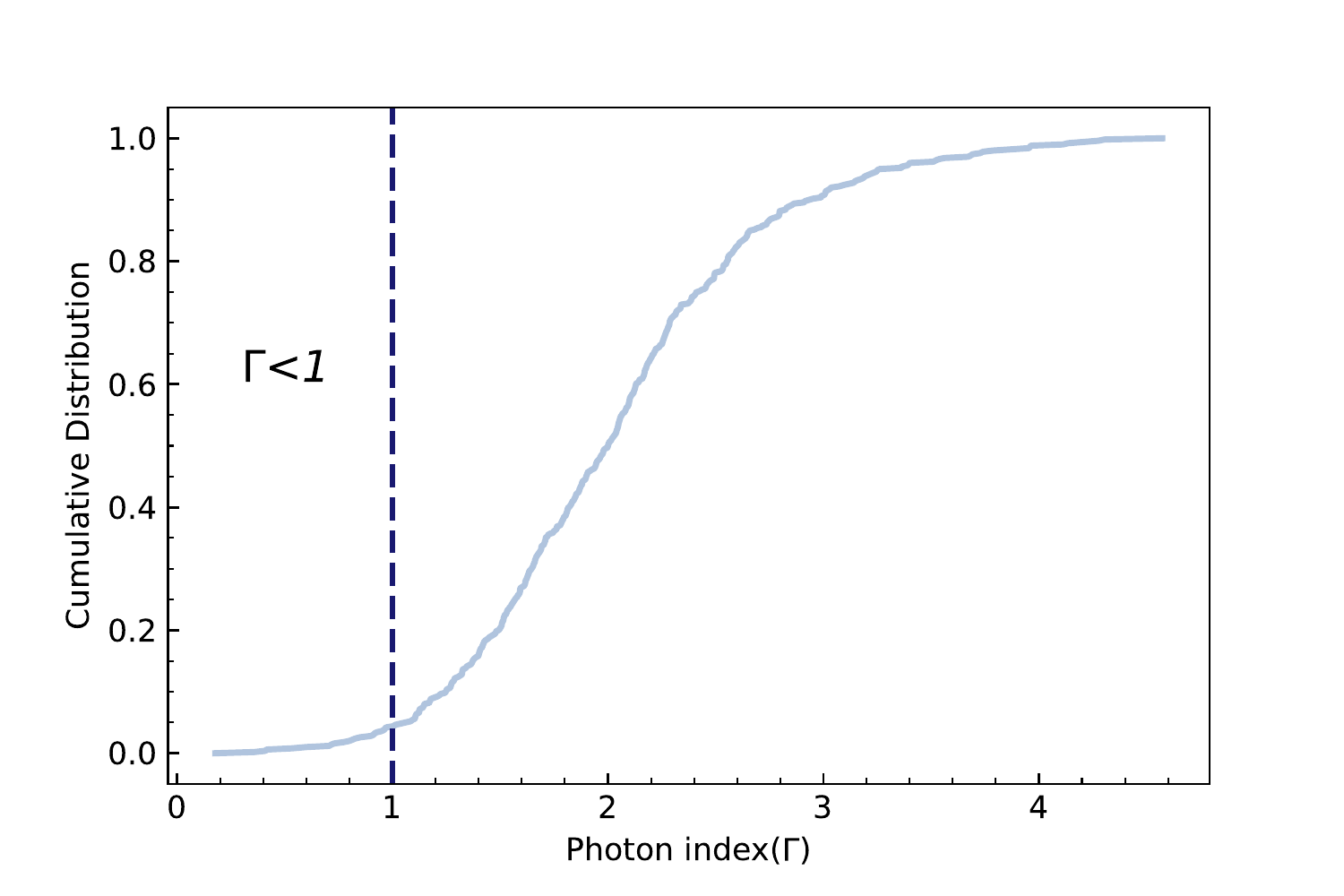}
\caption{Cumulative distribution of photon index over 500 simulated spectra with photon index equals 1.75. The vertical dashed line at $\Gamma =1$ indicates the probability of obtaining $\Gamma < 1$ which is $~4.4\%$.
\label{fig:3}}
\end{figure}

Unexpectedly, the first dataset observed at 8.9~d post-merger shows a marginally hard spectrum of $\Gamma = 1.04 ^{+0.44}_{-0.44}$ using our Bayesian analysis method (Fig.~\ref{fig:2}), in stark contrast to the softer spectrum inferred at a later time. This result is indeed consistent with previous reports using Cash statistic \citep{Margutti2018,Troja2018}. We therefore further test this result using Monte Carlo simulation.
We follow the {\tt Sherpa} analysis procedure to simulate 500 spectra with 
$\Gamma \approx 1.75$, which is the average of the later datasets (ObsIDs 20728-21371). The first two datasets have very low signal-to-noise ratio ($S/N$) due to low photon counts ($\sim 20$). Poisson noise was added to the simulated power law to achieve a similar $S/N$. Finally, we repeat the spectral analysis procedure aforementioned and obtain an average $\Gamma = 2.03\pm0.68$. The cumulative distribution of $\Gamma$ is shown in Fig.~\ref{fig:3} and the probability for obtaining $\Gamma <1$ is about 4.4\%.
The failure to reproduce $\Gamma \sim 1$ indicates that the observed hard spectrum is unlikely to be caused by statistical fluctuation in the {\it Chandra} observation. However, the large errors prevent us from drawing any firm conclusions regarding this peculiar spectral behavior. 

Furthermore, in addition to the standard toolkit {\tt Sherpa}, we perform the entire spectral analysis procedure using the Multi-Mission Maximum Likelihood framework \citep[\texttt{3ML},][]{Vianello2015}. We import from {\tt XPSEC} an absorbed power-law model with the neutral hydrogen column density fixed at the Galactic value, which is identical to the analysis using {\tt Sherpa}.
%With the slightly wider boundaries we set, uniform prior functions were adopted where normalization $K \sim \mathcal{U}({10^{-12},0.01})$ and $\Gamma \sim \mathcal{U}(-3.0,10.0)$. 
The {\tt 3ML} results are found to be in agreement with {\tt Sherpa}.
%the unchanged spectral shape as previously claimed.

We notice that the time-integrated $\Gamma$ obtained by other works (see Table~\ref{tab:results}) have similar values to the average of ours. This shows that intrinsic spectral evolution of individual time-resolved spectrum is non-negligible.
As a matter of fact, we also note that the errors on $\Gamma$ obtained using Cash statistic are asymmetric, indicating a significantly skewed parameter distribution, while our Bayesian posteriors are highly symmetric (see Fig.~\ref{fig:2} and Sect.~\ref{sec:diss} for further discussion).

\section{Discussion} \label{sec:diss}
%The hard-to-soft evolutionary trend of the observation episodes I, II, and III indicates that multiple emission epochs in the GRB afterglow have been going on. The hardening of the spectrum of episode IV could be a signal for another emission epoch. The Bayesian analysis performed in this work provides the first evidence of a possible, yet unexpected spectral change in the X-ray observations of GW170817/GRB170817A.

The X-ray emission in GRB afterglow is attributed to non-thermal synchrotron 
emission by relativistic electrons when the jet ejecta collide with the 
circumburst medium. In the external shock model, the synchrotron spectrum is 
divided into segments with different spectral slopes \citep{Sari1998,Granot2002}, 
and the breaks between the segments are where different characteristic frequencies 
located at. The relative positions of the cooling frequency $\nu_{\rm cool}$ 
and the minimum injection frequency $\nu_{\rm min}$ depends on the values of 
$\gamma_{\rm cool}$ and $\gamma_{\rm min}$ of the electron distribution. These values determine whether 
the so-called slow-cooling scenario ($\nu_{\rm min} < \nu_{\rm cool}$),
\begin{equation}
  F_{\nu{\rm ,slow}} \propto
  \begin{cases}
   	\nu^{2} & (\nu < \nu_{\rm a}), \\
    \nu^{1/3} & (\nu_{\rm a} < \nu < \nu_{\rm min}), \\
    \nu^{-(p-1)/2} & (\nu_{\rm min} < \nu < \nu_{\rm cool}), \\
    \nu^{-p/2} & (\nu > \nu_{\rm cool}),
  \end{cases}
  \label{eqn:slow}
\end{equation}
or fast-cooling scenario ($\nu_{\rm cool} < \nu_{\rm min}$),
\begin{equation}
  F_{\nu{\rm ,fast}} \propto
  \begin{cases}
    \nu^{2} & (\nu < \nu_{\rm a}), \\
    \nu^{1/3} & (\nu_{\rm a} < \nu < \nu_{\rm cool}), \\
    \nu^{-1/2} & (\nu_{\rm cool} < \nu < \nu_{\rm min}), \\
    \nu^{-p/2} & (\nu > \nu_{\rm min}),
  \end{cases}
  \label{eqn:fast}
\end{equation}
is taking place in the emission region.\footnote{The flux will diverge for $p \le 2$ when integrating the electron population and the synchrotron kernel \citep[see, e.g.,][]{Rybicki1979}.} 
%Below the absorption frequency $\nu_{\rm a}$, synchrotron self-absorption causes the spectral slope to harden to $\nu^2$. Above $\nu_{\rm a}$ is the characteristic slope for synchrotron emission, $\nu^{1/3}$. The highest energy slope always soften as $\nu^{-p/2}$. The slope of the segment between $\nu_{\rm min}$ and $\nu_{\rm cool}$ is either $\nu^{-(p-1)/2}$ if the cooling (radiation) timescale is longer than the dynamical timescale of the electrons, or fixed at $\nu^{-1/2}$ if the cooling timescale is shorter than the dynamical timescale.
The GRB afterglow is expected to evolve from fast to slow-cooling regime \citep[e.g.,][]{Sari1998,Granot2002}. 

According to Table~\ref{tab:results}, both {\tt Sherpa} and {\tt 3ML} successfully constrained $\Gamma$ that agrees within 1-$\sigma$ uncertainties. It can be seen that $\Gamma$ remains constant throughout the entire observing period within the error bars. Results derived from \texttt{Sherpa} and \texttt{3ML} present a consistent picture of a constant $\Gamma \sim 1.6$, in accordance with the interpretation that the $\nu^{-(p-1)/2}$ segment is being observed, assuming a typical electron power-law distribution index of $p = 2.2$ \citep{Alexander2017,Margutti2017,Margutti2018,Mooley2018,Troja2018}, consistent with the conventional slow-cooling scenario of GRB afterglows.

A possible scenario proposed by \citet{Piro2018} is that the appearance of an X-ray flare directly attribute to the increase of flux seen in episode III. X-ray flares are commonly seen in GRB afterglows \citep{Burrows2005,Zhang2006,Chincarini2007}, characterized by large flux variations in the X-ray light curves, and are mostly driven by the re-activation of the central source engines. In our analysis, the unabsorbed flux of the source reached its peak value $2.9 \times 10^{-14} ~\rm erg~\rm cm^{-2}~\rm s^{-1}$ at $t \sim 155$~d post-merger, and then decreased by a factor of $\sim 0.5$ within 6 days. \citet{Piro2018} estimated the time duration of this X-ray flare candidate was between 137 to 161 days post-merger, which is temporally consistent with our result. However, due to the lack of observations during the entire period of time, detailed comparison to other GRB X-ray flares is not possible.

 Placing the afterglow emission of GRB170817A in the slow-cooling regime provides constraints for the magnetic field. For instance, the cooling frequency can be written as $\nu_{\rm cool}= 3.7 \times 10^{14} E_{53}^{-1/2} n_{0}^{-1} (Y+1)^{-2} \epsilon_{{\rm B},-2}^{-3/2} T_{\rm d}^{-1/2} \, {\rm Hz}$, assuming on-axis and constant circumburst medium density \citep{Panaitescu2000}. Since we do not observe $\nu_{\rm cool}$ in the X-ray band, putting the isotropic energy $E \approx 5 \times 10^{46}$~erg \citep{Abbott2017}, the number density of the circumburst medium $n_0 \approx 5\times10^{-3}$~cm$^{-3}$ \citep[median value of short GRBs;][]{Fong2015}, the Compton parameter $Y = 0$, and the time in days $T_{\rm d} > 359$~d, it is estimated that $\epsilon_{\rm B} \sim 0.01$ assuming 10~keV for \textit{Chandra}'s upper energy limit, or $\epsilon_{\rm B} \sim 0.004$ if 100~keV is assumed. Although the estimation is derived from an on-axis model, it is consistent with the values from the cocoon model \citep[$\epsilon_{\rm B} \sim 0.01$, e.g., in][]{Mooley2018,Troja2018} and the off-axis model \citep[$\epsilon_{\rm B} \sim 0.001$, e.g., in][]{Troja2018}.

%\citet{Margutti2018} obtained an even earlier dataset at 2.34d which results in a non-detection in the X-rays. Since this observation is not yet public, we could not verify the non-detection using our Bayesian approach, and we note the reader that any explanation remains a speculation.

%Judging from the relative value of $\nu_{\rm min}$ and $\nu_{\rm cool}$, the slow-cooling scenario is preferred for the afterglow emission at 9 days, and $\nu_{\rm min}$ is below the {\it VLA} radio bands for all observations \citep{Alexander2017,Margutti2018}. Therefore, for $\nu_{\rm cool}$ to cross the {\it Chandra} X-ray band in $\sim$10-100 days requires $\epsilon_{\rm B}$ to be $\sim 2.1 \times 10^{-4}$ and decreases by a factor of 20 in about a week, assuming constant values for all other micro-physics parameters. We note that these values of $\epsilon_{\rm B}$ actually fall into the usual estimated range of $\sim 10^{-9}$-$10^{-4}$ for GRBs.

As a remark, \citet{Margutti2018} and \citet{Troja2018} also reanalyzed some of the {\it Chandra} datasets using Cash statistic. They reported similar spectra: 
$\Gamma = 0.95 ^{+0.95}_{-0.19}$ at $t=9$~d and $\Gamma = 1.6 ^{+1.5}_{-0.1}$ at $t=15$~d \citep{Margutti2018}; $\Gamma = 0.9 \pm 0.5$ at $t=9$~d and $\Gamma =1.6\pm 0.4$ at $t=15$~d \citep{Troja2018}. Comparing to our Gaussian-like 
posterior distribution of $\Gamma$ (Fig.~\ref{fig:2}), the highly asymmetric 
error bars might be an indication that the frequentist approach is 
limited by the poor $S/N$. Although these results are skewed towards 
harder values, they are consistent with our Bayesian results, possibly indicating a hard-to-soft evolution during the first episode. 

\section{Summary}\label{sec:con}
Taking the advantage of Bayesian statistics in dealing with the low-count X-ray data, we performed a detailed spectral analysis to all the public \textit{Chandra} observations of the binary neutron star coalescence GW170817/GRB170817A to date. 
The observed X-ray afterglow can be explained by a simple power-law spectral shape originated from slow-cooling synchrotron radiation. Throughout the one year period since the merger, the photon index is consistently shown to have remained constant at $\Gamma \sim 1.6$, suggesting an electron power-law distribution index $p = 2.2$. An unexpectedly hard spectrum of $\Gamma \sim 1$ observed at $\sim 9$~d post-merger is observed, though large uncertainties exist. The unabsorbed flux is found to be peaked at $t \sim 155$~d post-merger, which hints to a possible X-ray flare.
This is the first time Bayesian analysis is applied to the X-ray data of GRB170817A. With our straightforward derivation of the posterior distributions of spectral parameters and uncertainties which are consistent with other works, we have shown that the application of Bayesian inference is a viable method in analyzing the low $S/N$ data from this NS-NS merger as well as similar events.

%The observed evolution of the power-law spectral shape can be explained by the synchrotron slow-cooling scenario in the external shock model of GRB afterglow. We interpret the temporal evolving photon index, $1.5 < \Gamma <2$ , as the crossing of $\nu_{\rm cool}$ in the {\it Chandra} observing band. In the mean time, the unexpected hard spectra of $\Gamma \sim 1$ observed at $\sim 9$ days post-merger, though uncertainties exist, may indicate that we are witnessing the crossing of $\nu_{\rm min}$. 
%We therefore propose a possibly inhabited spectral change of GW170817/GRB170817A during its early stage of X-ray emission.

\acknowledgements
This project is supported by the Ministry of Science and Technology of the
Republic of China (Taiwan) through grants 105-2119-M-007-028-MY3,
106-2628-M-007-005 and 107-2628-M-007-003. HFY acknowledges support from the Swedish National 
Space Board and the Swedish Research Council (Vetenskapsr\r{a}det).

%\begin{thebibliography}{}

\bibliography{mybibfile}
%\end{thebibliography}

\end{document}